\documentstyle[12pt]{article}

\def\ord{{\cal O}}
\def\del{\delta}
\def\sa{\phantom{a}}

\begin{document}

\baselineskip 0.7cm

\begin{titlepage}
\begin{flushright}
UT-914
\\
November, 2000
\end{flushright}
\vskip 1.4 cm
\begin{center}
{\large{\bf
The Brown-Henneaux's central charge
from the path-integral boundary condition
}}
\vskip 1.2 cm
Hiroaki Terashima
\vskip 0.4 cm
{\it Department of Physics,University of Tokyo\\
Bunkyo-ku,Tokyo 113-0033,Japan}
\end{center}
\vskip 1.5 cm
\abstract{
We derive Brown-Henneaux's commutation
relation and central charge in the framework
of the path integral.
If we use the leading part of
the asymptotic symmetry to derive
the Ward-Takahashi identity,
we can see the central charge arises from
the fact that the boundary condition of the path
integral is not invariant under the transformation.
}
\end{titlepage}

\section{Introduction}
In the (2+1) dimensional spacetime with 
a negative cosmological constant $\Lambda=-1/l^2$,
Brown and Henneaux~\cite{BroHen86} have shown that
the asymptotic symmetry of the asymptotically $AdS_3$ spacetime
is the conformal group in 2 dimensions and
this symmetry is canonically realized by the Poisson bracket
algebra of the Hamiltonian generators with a central charge,
\begin{equation}
  c=\frac{3l}{2G}.
\label{central}
\end{equation}
This central charge was also obtained
in the Chern-Simons formulation of
the (2+1) dimensional gravity~\cite{BanadoCS}
or in the context AdS/CFT
correspondence~\cite{AdSCFT}.
Combining this central charge with the Cardy formula,
Strominger~\cite{Stromi98} has suggested that
the Bekenstein-Hawking entropy
of the BTZ black hole~\cite{BTZbh} can be understood as
the density of states of some conformal field theory.
For further generalizations,
see Refs.~\cite{Carlip99,NaOkSa00}.

In this paper,
we consider to derive this central charge
in the path integral formulation
since it was originally obtained by the canonical formulation.
In view of the equivalence of these
two approaches to quantum theory,
we must obtain the same result within the path integral.
In the context of the path integral,
the usual central charge is the quantum anomaly
which is understood as the Jacobian factor of
the path integral measure~\cite{Fujika87}.
However, Brown-Henneaux's central charge is
classical one because it exists at the level of
the Poisson bracket.
Thus, we want to clarify the origin of this
classical central charge
in the formulation of path integral.

\section{Transformations}
We consider the asymptotically $AdS_3$ spacetime
which is defined by the boundary condition~\cite{BroHen86},
\begin{eqnarray}
  g_{tt} =-\frac{r^2}{l^2}+\ord(1),
 & g_{tr} = \ord(1/r^3),
 & g_{t\phi} = \ord(1),   \nonumber \\
  g_{rr}  = \frac{l^2}{r^2}+\ord(1/r^4),
 & g_{r\phi} =\ord(1/r^3),
 & g_{\phi\phi} = r^2+\ord(1).
\label{adsbc}
\end{eqnarray}
The asymptotic symmetry of this spacetime
becomes
\begin{eqnarray}
 \xi^t &=&lT(t,\phi)+
   \frac{l^3}{r^2}\bar{T}(t,\phi)+\ord(1/r^4),\nonumber \\
 \xi^r &=&rR(t,\phi)+\ord(1/r), \nonumber \\
 \xi^\phi &=& \Phi(t,\phi)+
   \frac{l^2}{r^2}\bar{\Phi}(t,\phi)+\ord(1/r^4),
 \label{asymptotic}
\end{eqnarray}
where they satisfy
\begin{eqnarray}
 l\partial_tT(t,\phi) &=& 
     \partial_\phi \Phi(t,\phi)=-R(t,\phi),\nonumber \\
 l\partial_t\Phi(t,\phi)&=&
     \partial_\phi T(t,\phi),
\label{trphi}
\end{eqnarray}
and
\begin{eqnarray}
 \bar{T}(t,\phi)&=&
     -\frac{l}{2}\partial_tR(t,\phi),\nonumber \\
 \bar{\Phi}(t,\phi)&=&
      \frac{1}{2}\partial_\phi R(t,\phi).
\end{eqnarray}
This transformation preserves the above
boundary condition (\ref{adsbc})
and is the conformal group in 2 dimensions.

We may consider another transformation
which is the leading part of the asymptotic symmetry,
\begin{eqnarray}
 \xi'^t &=&lT(t,\phi),\nonumber \\
 \xi'^r &=&rR(t,\phi), \nonumber \\
 \xi'^\phi &=& \Phi(t,\phi), \label{leading}
\end{eqnarray}
where $T,R,\Phi$ again satisfy the above equations (\ref{trphi}).
Note that this transformation is {\em not} the asymptotic
symmetry since it breaks the boundary conditions for
$g_{tr}$ and $g_{r\phi}$.
However, the charge of this transformation is
the same as that of the asymptotic symmetry
as we see later.

\section{Action and Charge}
If we assume that the boundary of the spacetime is only
at infinity $r=r_\ast\to\infty$
whose unit normal vector is $u^a$,
the action becomes~\cite{Kterm}
\begin{equation}
 S = \frac{1}{16\pi G}\int_M \sqrt{-g}\;\left(R-2\Lambda\right) \;d^3x+
   \frac{1}{8\pi G}\int_{r=r_\ast}\sqrt{-\gamma}\;\Theta \;d^2x,
\end{equation}
where $\gamma_{ab}$ is the induced metric on
the boundary $r=r_\ast$ defined by
$\gamma_{ab}=g_{ab}-u_a u_b$
and $\Theta^{ab}$ is the extrinsic curvature of the boundary
defined by $\Theta^{ab}=\gamma^{ac}\nabla_c u^b$.
The generic variation, namely
$\del g_{ab}\neq0$ at $r=r_\ast$, of this action is~\cite{York86,BroYor93}
\begin{equation}
 \del S=-\frac{1}{16\pi G}\int_M \sqrt{-g}\;\tilde{G}^{ab}\;
    \del g_{ab}\;d^3x
  -\frac{1}{16\pi G}\int_{r=r_\ast}\sqrt{-\gamma}\;
     \Pi^{ab}\;\del\gamma_{ab}\;d^2x,
\end{equation}
where $\tilde{G}^{ab}=R^{ab}-\frac{1}{2}g^{ab}R+\Lambda g^{ab}$
and $\Pi^{ab}=\Theta^{ab} - \Theta\gamma^{ab}$.
By using this formula,
we find that the change of the action under
the transformation
$\del g_{ab}=\nabla_a\zeta_b+\nabla_b\zeta_a$ becomes
\begin{eqnarray}
 \del_{\zeta} S &=& 
 -\frac{1}{8\pi G}\int_M \sqrt{-g}\;\tilde{G}^{ab}\;\nabla_a\zeta_b\;d^3x
  \nonumber \\
  & &
{}-\frac{1}{8\pi G}\int_{r=r_\ast}\sqrt{-\gamma}\;
  \left[ \Pi^a_{\sa b}\;{\cal D}_a \tilde{\zeta}^b +
   \eta\,\left(\Theta^{ab}\,\Theta_{ab}-\Theta^2\right)\right]\;d^2x,
\end{eqnarray}
where $\eta=\zeta^a u_a$ and $\tilde{\zeta}^a=\zeta^a-\eta \,u^a$
is the tangential part of $\zeta^a$ to the boundary $r=r_\ast$.
Unfortunately, this would diverge in the limit of
$r_\ast\to\infty$.
Therefore, it is usual to subtract a functional of the boundary
data $\gamma_{ab}$ from the action~\cite{BroYor93}.
We here choose so that
\begin{eqnarray}
 \del_{\zeta} S &=& 
 -\frac{1}{8\pi G}\int_M \sqrt{-g}\;\tilde{G}^{ab}\;\nabla_a\zeta_b\;d^3x
  \nonumber \\
  & &
{}-\frac{1}{8\pi G}\int_{r=r_\ast}\sqrt{-\gamma}\,
  \left[ \left(\Pi^a_{\sa b}-\hat{\Pi}^a_{\sa b}\right)
   \,{\cal D}_a \tilde{\zeta}^b \right.
 \nonumber \\
  & &
  \left.\qquad\quad+\eta\left(\Theta^{ab}\,\Theta_{ab}-\Theta^2-
     \hat{\Theta}^{ab}\,\hat{\Theta}_{ab}+\hat{\Theta}^2\right)\right]\;d^2x,
\end{eqnarray}
where the hats mean that they are evaluated
by the $M=J=0$ BTZ black hole
rather than $AdS_3$ spacetime for a technical reason.

We can identify the charge of Brown and Henneaux as
\begin{equation}
  J[\xi]=-\frac{1}{8\pi G}\lim_{r_\ast\to\infty}
  \int_{r=r_\ast}d\phi\sqrt{\sigma}
  \left(\Pi^a_{\sa b}-\hat{\Pi}^a_{\sa b}\right)\tilde{\xi}^bn_a,
 \label{charge}
\end{equation}
where $n^a$ is the unit normal vector of the time slice and
$\sigma_{ab}$ is the induced metric on the boundary $r=r_\ast$
of the time slice.
To our knowledge, this alternative expression (\ref{charge})
for Brown-Henneaux's charge has not been discussed before.
It is easy to check that this charge is actually
identical to Brown-Henneaux's charge
by expanding around the $M=J=0$ BTZ black hole.
See, however, Ref.~\cite{BroYor93} for a related definition
of the global charge.
Note that $J[\xi]=J[\xi']$ for the transformations
in Eqs.~(\ref{asymptotic}) and (\ref{leading})
since the non-leading terms
does not contribute to the charge.
After a straightforward calculation, we find that
the change of this charge under the asymptotic symmetry
becomes
\begin{equation}
\del_{\xi_2} J[\xi_1]=J\Bigl[[\xi_1,\xi_2]\Bigr]+K[\xi_1,\xi_2]+\cdots,
\label{delj2}
\end{equation}
where $K[\xi_1,\xi_2]$ is Brown-Henneaux's central charge,
\begin{equation}
  K[\xi_1,\xi_2]= -\frac{1}{8\pi G}\int d\phi
  \left(T_1\partial_\phi^3+\Phi_1l^3\partial_t^3\right)l \Phi_2,
\end{equation}
and `$\cdots$'' means the terms which vanish
by using the equations of motion.
By Fourier transformation, this becomes the usual central term
with the central charge (\ref{central}).
This provides an alternative derivation of
Brown-Henneaux's central charge which is 
simpler than the original one.
On the other hand, an interesting aspect of
the leading transformation (\ref{leading}),
which motivated the present work,
is that the change of the charge (\ref{charge}) gives rise to
\begin{equation}
\del_{\xi'_2} J[\xi_1]=J\Bigl[[\xi_1,\xi_2]\Bigr]+\cdots,
\label{delj}
\end{equation}
without any central charge.
Since the remaining quantities which appear in
the Ward-Takahashi identity
are the same as those of the asymptotic symmetry,
one might think that we can obtain the commutator
without any central charge if we use this transformation.
We thus want to understand the origin of the central charge
by this transformation.

\section{Commutation relation}
We begin with the path integral
\begin{equation}
  \left\langle J[\xi_1] \right\rangle=\int_B d\mu\,J[\xi_1]\, e^{iS},
\end{equation}
where $d\mu$ and $B$ denote the measure and boundary condition
of the path integral, respectively.
To obtain the Ward-Takahashi identity,
we perform the infinitesimal change of the integration variable
corresponding to the leading transformation $\xi_2'$
rather than the asymptotic symmetry.
(We assume that $d\mu$ is invariant under this
transformation since we want to calculate the classical
central charge.)
We then have the Ward-Takahashi identity,
\begin{equation}
 \left\langle \del_{\xi'_2} J[\xi_1] \right\rangle
 =-i \left\langle {\rm T}^\ast\,J[\xi_1]\;\del_{\xi'_2}S\right\rangle
  -\Delta[\xi'_1,\xi'_2],
\end{equation}
where
\begin{equation}
 \Delta[\xi'_1,\xi'_2] \equiv 
   \left( \int_{B+\del_{\xi'_2}B} -
      \int_B\right) d\mu\, J[\xi_1] \, e^{iS},
\label{delta}
\end{equation}
and the boundary condition $B+\del_{\xi'_2}B$ denotes
that the transformed metric $g_{ab}+\del_{\xi'_2}g_{ab}$ must
satisfy the asymptotically $AdS_3$ condition (\ref{adsbc}).
Note that we have an extra term $\Delta[\xi'_1,\xi'_2]$
because the leading transformation breaks the boundary
condition of the path integral $B$.
By using the asymptotically $AdS_3$ condition (\ref{adsbc}),
we evaluate the first term of the right-hand side.
Then, one finds that
\begin{eqnarray}
  \left\langle {\rm T}^\ast\,  J[\xi_1]\; \del_{\xi'_2}S  \right\rangle
    &=& -\left\langle {\rm T}^\ast\,  J[\xi_1]\;
           \int dt \;\partial_t J[\xi_2] +\cdots   \right\rangle \nonumber \\
    &=& -\int dt \;\partial_t \left\langle {\rm T}^\ast\,  J[\xi_1]\;
            J[\xi_2] +\cdots   \right\rangle \nonumber \\
    &=& -\int dt \;\partial_t \left\langle {\rm T}\,  J[\xi_1]\;
            J[\xi_2] +\cdots   \right\rangle \nonumber \\
    &=& \left\langle \Bigl[ J[\xi_1],J[\xi_2 ] \Bigr]+\cdots\right\rangle,
\end{eqnarray}
where ``$\cdots$'' again means the terms which vanish
by using the equations of motion.
Here we used the standard Bjorken-Johnson-Low
argument to convert the T$^\ast$-product to
the canonical T-product.
Combining with Eq.~(\ref{delj}),
we can obtain the anomalous commutator of two charges as
\begin{equation}
   \left\langle \Bigl[ J[\xi_1],J[\xi_2 ] \Bigr] \right\rangle= 
   \left\langle i J\Bigl[[\xi_1,\xi_2]\Bigr]\right\rangle+
     i\Delta[\xi'_1,\xi'_2],
\end{equation}
by using the equations of motion.

Next, we want to evaluate the anomalous
term $\Delta[\xi'_1,\xi'_2]$
which is defined by Eq.~(\ref{delta}).
By performing the infinitesimal change of
the integration variable corresponding to
the inverse transformation of $\xi_2'$ in the first integral
and that of $\xi_2$ in the second integral of Eq.~(\ref{delta}),
the integrals become
\begin{eqnarray}
  \int_{B+\del_{\xi'_2}B}d\mu\, J[\xi_1] \, e^{iS}
   &=& \int_B d\mu\,\left(J[\xi_1]-iJ[\xi_1]\;\del_{\xi'_2}S
                -\del_{\xi'_2} J[\xi_1] \right) \, e^{iS}, 
     \nonumber \\
  \int_B d\mu\,J[\xi_1] \, e^{iS}
   &=& \int_B d\mu\,\left(J[\xi_1]-iJ[\xi_1]\;\del_{\xi_2}S
                -\del_{\xi_2} J[\xi_1] \right) \, e^{iS}.
\end{eqnarray}
Note that the boundary condition of both of the path integral
become the same.
Since $\del_{\xi_2}S=\del_{\xi'_2}S$ but
$\del_{\xi_2} J[\xi_1]\neq\del_{\xi'_2} J[\xi_1]$
as Eqs.~(\ref{delj2}) and (\ref{delj}) show,
one can obtain that
\begin{equation}
  \Delta[\xi'_1,\xi'_2] = \left\langle 
      \del_{\xi_2} J[\xi_1]-\del_{\xi'_2} J[\xi_1]
      \right\rangle
   = \left\langle K[\xi_1,\xi_2] \right\rangle,
\end{equation}
by using the equations of motion.
Thus, we finally obtain
\begin{equation}
   \left\langle \Bigl[ J[\xi'_1],J[\xi'_2 ] \Bigr] \right\rangle= 
   \left\langle i J\Bigl[[\xi'_1,\xi'_2]\Bigr]\right\rangle+
     iK[\xi_1,\xi_2],
\end{equation}
which is consistent with
the result of Brown and Henneaux.

\section{Discussion}
We have reproduced Brown-Henneaux's commutator and
central charge in the frame of the path integral.
By using the leading transformation (\ref{leading})
of the asymptotic symmetry
to derive the Ward-Takahashi identity,
we have shown that Brown-Henneaux's central charge
arises from the path integral boundary condition.
That is, the central charge arises from the fact
that {\em the boundary condition}
of the path integral is not invariant under the transformation.
This is in contrast to the usual quantum case,
where the anomaly arises from the fact that
{\em the measure} of the path integral is not invariant
under the relevant transformation.
Other classical central charges,
such as in $N=2$ supersymmetric
theory~\cite{FujOku98},
may also be understood as above
in the path integral formalism.

Of course, we can derive the above commutator by using
the asymptotic symmetry itself.
Then, the central charge arises from the transformation
law of the charge.
However, the present analysis suggests the possibility
that the classical central charge may arise
in more general theories if the boundary condition
of the path integral is non-trivial.
A path integral generally provides a more
transparent framework to study various topological
properties such as related to the black hole.
In view of this,
it is important to understand the origin of the
central charge in path integral.
It is also gratifying that one can derive
the fundamental result in a variety of ways.

\section*{Acknowledgments}
The author thanks K. Fujikawa for
helpful discussions and comments.

\end{document}